**Title Page**

**Qualifying threshold of "take-off" stage for successfully disseminated creative ideas**

Guoqiang Liang[1,2], Haiyan Hou[1], Xiaodan Lou[3], Zhigang Hu[1]

*(1. WISE Lab, Dalian University of Technology, Dalian, Liaoning 116023, China*

*2. School of Informatics, Computing, and Engineering, Indiana University, Bloomington, IN 47408, U.S.A.*

*3. Institute of Systems Science, Beijing Normal University, Beijing, 100875, China)*

**ORCID of the authors**

Guoqiang Liang: 0000-0002-9669-4048
Haiyan Hou: 0000-0002-2790-9973
Zhigang Hu: 0000-0003-1835-4264

**Corresponding author:**

Zhigang Hu, WISE_Lab, Dalian University of Technology, Dalian 116023, Liaoning, China

E-mail: huzhigang@dlut.edu.cn, Tel: +8613591791279.

**Abstract**

The creative process is essentially Darwinian and only a small proportion of creative ideas are selected for further development. However, the threshold that identifies this small fraction of successfully disseminated creative ideas at their early stage has not been thoroughly analyzed through the lens of Rogers's innovation diffusion theory. Here, we take highly cited (top 1%) research papers as an example of the most successfully disseminated creative ideas and explore the time it takes and citations it receives at their "take-off" stage, which play a crucial role in the dissemination of creativity. Results show the majority of highly cited papers will reach 10% and 25% of their total citations within two years and four years, respectively. Interestingly, our results also present a minimal number of articles that attract their first citation before publication. As for the discipline, number of references, and Price index, we find a significant difference exists: Clinical, Pre-Clinical & Health and Life Sciences are the first two disciplines to reach the $C_{10\%}$ and $C_{25\%}$ in a shorter amount of time. Highly cited papers with limited references usually take more time to reach 10% and 25% of their total citations. In addition, highly cited papers will attract citations rapidly when they cite more recent references. These results provide insights into the timespan and citations for a research paper to become highly cited at the "take-off" stage in its diffusion process, as well as the factors that may influence it.

**Introduction**

The creative process is essentially Darwinian (Simonton, 1997). That is, creativity entails some variation-selection process that generates numerous conceptual combinations. More specifically, in the idea creation process, one constructs ideational combinations in a spontaneous, unpredictable way. Individuals have no a priori way of foreseeing which ideational combinations will prove most fruitful. Only a small proportion of these combinations is then selected for further development. In academia, creative ideas are contained in written documents constituting academic papers. Researchers build their work on the combinations of existing knowledge and submit them to related journals where they are subjected to selection along with hundreds of manuscripts of the same genre submitted by researchers with similar aspirations. Only a small set then survives the selection process (Boyack and Klavans 2014; Simonton 1997; Uzzi et al. 2013). Nor does the selection process end here because not all scientific publications have the same impact on the scientific community, as judged by citation indices (Fortunato et al. 2018; Roth et al. 2012; Yu et al. 2014). Often only a tiny fraction of the publications managed to be successful among their contemporaries (Simonton 1997).

However, what features distinguish this tiny fraction of publications that are successful? Is there a threshold of minimal citation count during the early period, after an article's publication, that is required for its future success? This is of high significance for the evaluation of newly published articles within a short time window, as citation counts are attractive raw data for scientific impact evaluation and reward allocation (Bourdieu 1991; R. K. Merton 1968), and a good proxy for scientific creativity (Bornmann and Daniel 2008; Lee et al. 2015). Early studies from bibliometrics and complex networks have examined article-, author-, journal-, and field-related factors that are connected to citation impact (Beaudry and Larivière 2016; Didegah and Thelwall 2013b; Haslam et al. 2008; Onodera and Yoshikane 2015; Roth et al. 2012; Tahamtan et al. 2016; D. Wang et al. 2013; Yu et al. 2014). Some studies have also focused on the citation distribution based on the life-cycle theory (Bouabid 2011; Min et al. 2018), as well as graph mining (Pobiedina and Ichise 2016). However, the diffusion of scientific citations remains relatively less explored (Min et al. 2018).

Diffusion is the process by which an innovation is communicated through certain channels over time among the members of a social system (Rogers 1995). Most innovations have a S-shaped rate of adoption and there are variations in the slope of the "S" from innovation to innovation, which indicates the rate of diffusion. According to Rogers, after slow initial development, the S-shaped diffusion curve usually "takes-off" at about a 10-25 percent adoption rate. This period acts as an accelerant in the diffusion process of these creative ideas and plays a crucial role in the dissemination within its research field and to other areas (Rogers 1995). In fact, the citation diffusion of scientific publications fits the Bass diffusion model, which is based on and extended from Rogers diffusion theory (Min et al. 2018). Innovating upon prior studies, here, we explore the threshold at the "take-off" stage of the most successfully disseminated ideas in science, which is represented by highly cited papers. Getting a high-quality article adopted by the scientific community, even when it has obvious advantages, is often difficult (Hu and Wu 2018; Li and Ye 2016; van Raan 2004). Many of these articles require a lengthy period, often many years from the time they were accepted by a journal to the time they are widely adopted by contemporaries. Therefore, the study of "take-off" stage on highly cited papers will shed light on the funding agency for timely assessment of funded research, facilitate the early recognition of academic performance, and illuminate research fronts detection.

This article is outlined as follows. We first discuss work related to our study, giving attention to prior studies on key factors and prediction methods of citation impact, as well as Rogers's innovation diffusion theory. Then we detail the data and indicators for our analysis. Next, we present our findings concerning the time period (T) and citations (C) of the earliest 10% and 25% of total citations per article. Following, we study factors that may affect the results, such as disciplines, number of references, and Price index. Finally, we discuss the results and conclude with a summary of our findings, limitations, and thoughts for future research.

**Related work**

*Determiners of citation impact: a short review of the literature*
The citation distribution in science is highly skewed so the majority of articles are scarcely cited while a tiny fraction of others is highly cited (Tahamtan et al. 2016). Many researchers have investigated the features that make some papers cited more than others, from author-, article-, journal-, and field-related factors that are connected to citation impact.

Studies on author-related factors have shown that gender inequality exists in citing behavior, because publications of female researchers are less cited than are those of male researchers, which may be a result of a cumulative advantage effect of male researchers tending to have a higher relative publication output, based on a large-scale study of 8,500 Norwegian researchers and more than 37,000 publications covering all areas of knowledge (Aksnes et al. 2011). Additionally, the fame of an author's name is also a predictive factor of future citations for articles, as shown by a study of articles published in 1996 of three primary journals in social-personality psychology (Haslam et al. 2008), and later verified by analysis of publications after 5 and 10 years on the real-world dataset extracted from AMiner (Yan et al. 2011b). Author collaboration, that is, the number of authors of an article, also has a general influence on citation counts (Bornmann and Daniel 2006; Fortunato et al. 2018; Kong et al. 2017; Lee et al. 2015).

In addition, article-related factors contribute to citations, such as number, impact, recency of references, age of the paper, and document type. After examining different factors affecting the number of citations across subject field, Onodera and Yoshikane (Onodera and Yoshikane, 2015) found that Price index and number of references were the first two predictors of citations as derived through negative binomial multiple regression. This conflicts with the conclusion drawn from data about the nanotechnology field between 2007 to 2009 (Didegah and Thelwall 2013a), which states impact of the cited references and impact factor of the publishing journal are the most effective determinants of citation counts. Researchers also tend to show more interest in recent publications because the value of knowledge decreases as times passes and, thus, recent published papers are more likely to attract citations (Bornmann 2013; Egghe et al. 1995; Gosnell 1943; Kong et al. 2017; R. K. Merton 1961). Several studies have investigated the connection between document type and citations and found that review papers achieve more citations in general (Tahamtan et al. 2016).

It is traditional wisdom that high impact journals attract more attention, therefore, papers published in these journals have a higher potential of being cited than those in other journals. After investigating publications during 2006-2007 by staff of the School of Environmental Science and Management at Southern Cross University, Vanclay (Vanclay 2013) revealed that journal impact factor and type will affect the citations of an article and suggested that writing substantial review articles, then submitting to high impact journals, will increase the potential citations in the future, although some researchers have held an opposing viewpoint in regards to this finding (Roldan-Valadez and Rios 2015). Additionally, articles in relatively small research fields generally receive fewer citations than those in fields with many publications (King 1987), and the field and time range of scientific publications can reflect the research fronts of this field. Researchers can gain insight into the papers' impact at the research front by looking at the number of times papers are cited in this field (H.F. Moed et al. 1985).

*Prediction methods of citation impact*

Trying to track all publications to build our work would not be viable because of the exponential growth of scientific literature (Mukherjee et al. 2017). Many researchers have developed different approaches to predict an article's future citations and forecast which kind of literature is more likely to attract scientists' responses. Some researchers have regarded citation count prediction as link prediction from a complex network viewpoint. Zhou et al. (Zhou et al. 2018) focused on h-Salton and h-AA link prediction methods, which are variants of the Salton and AA indices, to measure the importance of nodes in citation networks. They argued that h-index-based indicators have a positive effect on the application of link prediction methods and are suitable for measuring the importance of citation networks. Pobiedina and Ichise (Pobiedina and Ichise 2016) suggested a new feature based on frequent graph pattern mining to improve the accuracy of citation count prediction from a network point of view. Roth et al. (Roth et al. 2012) studied papers published between 1970 and 1999 on their citation network structures. They explored the relationship between references and quality of citing paper and found that papers with more references are more likely to get cited within the field. Papers with a low reference recency are more likely to be more cited in physics.

Some researchers have proposed different models to predict citations. For instance, Callaham et al. (Callaham et al. 2002) conducted a non-parametric modeling technique of regression trees to predict the future citations of 204 publications submitted to a 1991 emergency medicine specialty meeting. Their results showed that journal impact factor is the strongest predictor of citations per year, followed by the newsworthiness and subjective quality of the article. Similarly, Yan et al. (Yan et al. 2011a) utilized several features of highly cited papers, such as topic rank and diversity, article age, H-index, fame of the authors as input, and use of linear regression (LR), k-nearest neighbor (kNN), support vector regression (SVR), and classification and regression tree (CART) as the predictive models. They found that the prediction after a longer period can achieve the best accuracy with models of CART. They also found that authors have bias in citing references, as author's fame and publication venue are the most predictive features of future citations. Wang et al. (Wang et al. 2013) derived a WSB model to predict the long-term citations of individual papers that includes, in addition to the cumulative advantage and aging, a fitness parameter that accounts for the perceived novelty or importance of scientific publications. Their model has quickly drawn attention after publication because the WSB model correctly approximated the citation range for 93.5% of papers 25 years later. Yu et al. (Yu et al. 2014) regarded the features of author, journal, citation, article type, and publication date as a paper's feature space, which they constructed through stepwise multiple regression analysis. They argued that the constructed paper's feature space is significant at the 0.01 level to predict the citation impact of a paper in Information Science & Library Science field after the first 5 years of publication.

Others have studied this question from their citation distributions of scientific articles. For example, Bouabid H (Bouabid, 2011) assumed that the citation distribution of publications is like an inverse U-shaped curve and classified two main stages in an article's life-cycle which is split by the citations peak and proposed a model to represent the naturally observed citations distribution and citations aging in a diachronous approach. They then explored the time that an article will continue being cited and the rate of decline, and found that the life-cycle of publications is infinite when the residual citations are greater than or equal to zero. Min et al. (Min et al. 2018) regarded the citation process as an innovation diffusion process and studied the citation dynamics based on Bass diffusion model. They suggested a saturation level to roughly estimate an article's current stage in its citation life cycle and the potential of future citations. However, the features that affect citation counts are still poorly understand from an innovation diffusion point of view.

*Innovation diffusion theory*
Getting a new idea adopted is often difficult even when it has obvious advantages. Many innovations, ideas, products, technologies, or services require a lengthy period to be widely accepted. Innovation related factors (such as relative advantage, complexity, trialability, observability, compatibility), diffusion channels (such as mass media channels, interpersonal channels), time (such as rate of adoption, innovation-decision process), and social system (such as social norms, structure of a social system, opinion leaders) are main elements that affect the diffusion process. The diffusion process of these innovations, proposed by Rogers in the 1960s, seeks to understand this process and offer a guidance in marketing, management science, and public policies (Rogers 1995).

Innovation, as defined by Rogers (Rogers 1995), is an idea, practice, or object that is perceived as new by an individual or other unit of adoption. It does not matter whether or not the idea is objectively new, but whether the individual perceives it as such. That is, if the idea seems new to the adopter, it is an innovation, otherwise, it is not. Moreover, the innovation can either involve new knowledge or existing knowledge, as someone may have known about the innovation for some time but may not have responded (such as favorable or unfavorable, adopt or reject) to it. Diffusion is the process by which an innovation is communicated through certain channels over time among the members of a social system (Rogers 1995). As time is one of the main elements of the diffusion process and a key concern in this article, here we review the time dimension of Rogers theory in a more detailed way.

The function of time is involved in three aspects: (1) The innovation-decision process by which an individual becomes aware of the innovation, forms an attitude toward it, and makes a choice to adopt or reject this innovation. Knowledge, persuasion, decision, implementation, and confirmation are the five main steps in this process. (2) The period of an innovation's adoption cycle, during which an individual adopts it, is what distinguishes these individuals as "innovators", "earlier adopters", "earlier majority", "late majority", and "laggards." (3) The rate of adoption, which means the relative speed of an innovation that is adopted by a unit of adoption or members within the social system. Only a few innovative individuals adopt the innovation at first, later, more and more individuals adopt in each succeeding time period and, eventually, the adoption rate slows. When we plot the accumulated number of individuals adopting the innovation over time, it results an S-shaped curve (Figure 1). However, the slope of the S-shaped curve is different from innovation to innovation, as some innovation diffuse in a relatively rapid approach and some at a slower rate. Significantly, for successful innovations, the S-shaped diffusion curve usually takes off at about 10-25% of accumulative frequency.

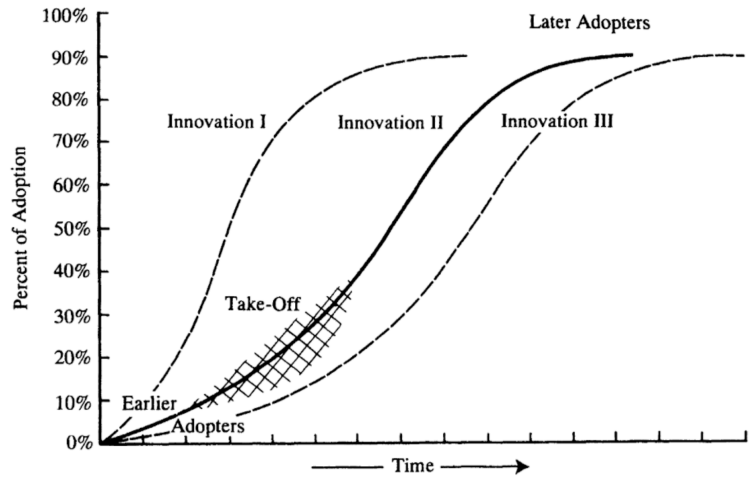
Fig. 1 Rogers Innovation diffusion model (Rogers 1995)

*Citation-based innovation diffusion*

Innovation diffusion is a theory that seeks to explain how, why, and at what rate a new idea, practice, or object spreads over time among the members of a social system. In academia, a new idea is documented in scientific publications. Being accepted for publication indicates the new ideas were recognized by reviewers and being cited after publication further shows its influence on the scientific community. In this sense, the citation behavior between scientific publications can be viewed as a scientific impact diffusion process, where the number of citations represent impact depth of diffusion in scientific community. Based on the assumption that researchers cite the works that influence them, some studies have used citation as a proxy for innovation diffusion. Zhai et al. (Zhai et al. 2018) used Latent Dirichlet Allocation (LDA) citations extracted from Scopus between 2003 and 2015 to measure the diffusion of innovation in different stages and subjects. The results showed that as LDA is transferred into different areas, and the adoption of each subject was relatively adjacent to those with similar research interests. Min et al. (Min et al. 2018) tried to understand the diffusion mechanisms underlying the citation process. Their data set was based on essays of 629 Nobel Prize winners in Chemistry, Physics, Physiology or Medicine, and Economic Science published from 1900 to 2000 and citation data until 2011 indexed in Web of Science. Using Bass model, they quantified and illustrated specific diffusion mechanisms which have been proven to exert a significant impact on the growth and decay of citation counts.

Indeed, citations cannot totally represent the adoption of innovation and researchers have pointed to a number of concerns about citations analysis, including misleading or wrong citations and complex citing motives (Bornmann and Daniel 2008). However, citation-based data still has much more advantage for innovation diffusion research compared to the data collected from questionnaires, interviews, and in-depth case studies because, on one hand, citations are easily accessible in electronic form for revealing the content of the diffusion process, so we can map the entire diffusion process of the innovation, especially in the big-data era. On the other hand, by using the citation data, researchers can easily track the historical roots of scientific ideas as well as forecasting the research trends.

**Methodology**

*Data*

The empirical analysis in this article is based on a dataset of highly cited research articles that were published in 2008 by Web of Science (WoS), which defined highly cited papers as articles that received the top 1% of citations within the same research field and publication year based on InCites (http://incites.thomsonreuters.com/) until July/August 2018. We also obtained their citing articles until the year 2017. As a result, 7927 highly cited articles and 1,742,036 citing articles (10 years citation window) were obtained in total. Studies have shown that a short citation window (usually 1 or 2 years) has bias for at least two reasons (Wang 2013): (1) There is a difference in the recognition time period at the field level, such as the mathematics and social sciences fields need much longer time to be recognized than the biomedical field. Thus, a short citation window will benefit fields that have a short recognition time. (2) The aging pattern of different papers varies. Some papers have shown "delayed rise-slow decline", and some have shown "early rise-rapid decline." Therefore, a short time window underestimates papers that are more valuable and influential. On account of the bias that may be caused by the short citation time window and the earliest highly cited papers we can obtain in WoS is 2008, hence, we choose the highly cited papers published in 2008 as our dataset.

*Indicators*

According to Rogers innovation diffusion theory, the diffusion process usually takes off at 10-25% of accumulative frequency (Rogers 1995). Therefore, we explore the time and related citation counts when they reach the 10% and 25% of total citation counts, respectively. There are four indicators employed in this study:

$T_{10\%}$ and $C_{10\%}$ mean the time and related citation counts of the 10$^{th}$ percent of accumulative frequency, respectively. These two indicators are responsible for the lower threshold of takes off stage.

$T_{25\%}$ and $C_{25\%}$, similar to the above two indicators, indicate the time and corresponding citations of the 25$^{th}$ percent accumulative frequency, respectively. These two indicators are measurements for the upper threshold of the "take-off" period.

*Number of references and Price index classification methods*

We classify the references list into three groups:
(1) Limited knowledge broadness: the number of references in each paper is equal to or less than 21.
(2) Moderate knowledge broadness: the number of references in each paper is between 22 and 41.
(3) Large knowledge broadness: the number of references in each paper is equal to or higher than 42.

In addition, we classify the Price indices into three groups, like we do with the classification of number of references:
(1) Older knowledge roots: the ratio of references to literature published in the last five years is less than or equal to 0.43.
(2) Moderate knowledge roots: the ratio of references to literature published in the last five years is between 0.44 and 0.71.
(3) Newer knowledge roots: the ratio of references to literature published in the last five years is equal to or greater than 0.72.

Knowledge broadness and recency limits are determined by the percentile values in the distribution of references by number ($P_{25\%} = 21$; $P_{75\%} = 41$), and Price index ($P_{25\%} = 0.43$; $P_{75\%} = 0.71$). Note that the "limited-", "moderate-", and "large-" knowledge broadness, or "older," "moderate," and "newer" knowledge roots, should be understood in relative terms. In this sense, one could argue that 50 references is not very many; but relative to this study, this amount is high, and belongs to the highest classification in number of references. The purpose of categorizing number of references and Price indices in a three-class category is to compare differences along three distinct stages in the broadness and recency of scientists' referencing practice.

**Results**

*"Take-off" stage of highly cited papers*

Figure 2 presents the S-shaped curves of highly cited papers as well as an average curve of these curves based on Rogers's innovation diffusion theory. The horizontal axis presents the time (year) from the first citation to their total citations until the year 2017. The vertical axis gives the accumulated percentage of the total citation counts for each paper. Papers with an accumulated percentage between 10% and 25% are regarded as the "take-off" period of successfully diffused ideas, which is the main focus of this study.

We observe that 87.18% (6911) articles reach the 10% of accumulated citations two years after their publication and the average citation count is 50.46. The percentage of articles that reach 25% of their total citations four years after publication is 96.58% (7656) and their mean number of citations is 102.93. This means the large majority of highly cited papers will reach 10% of their total citations within two years and 25% within four years. In addition, we find that there is still a minimal number of papers showing the "delayed rise" pattern. Over all, the average time is 1.78 and 3.05 years for these articles to reach 10% (mean value = 51.79) and 25% (mean value = 104.07) of their total citations. Interestingly, our results also show that there are 0.66% (52) of articles that accumulate their first citations before publication.

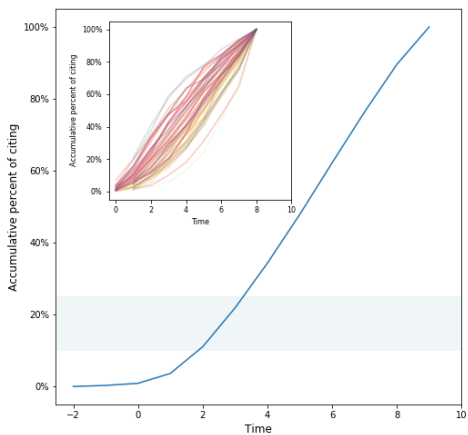

Fig. 2 S-shaped curves of highly cited papers. The solid blue line indicates an average S-shaped curve of highly cited papers that fitted by moving average method, and the shade indicates the "take-off" stage of curves.

*Disciplinary differences*

In this section, we compare the time it takes and citations it receives in the "take-off" stage among disciplines. To decide the papers' discipline, we based our classification on the GIPP Mapping Table (http://ipscience-help.thomsonreuters.com/inCites2Live/indicatorsGroup/aboutHandbook/appendix/mappingTable.html). In addition, the discipline of an article is measured by full counts which means for a paper belonging to two or more disciplines is assigned to each discipline. As a result, there are 2062, 2458, 2724, 1770, 622, and 10 papers in Clinical, Pre-Clinical & Health, Life Sciences, Physical Sciences, Engineering & Technology, Social Sciences, and Arts & Humanities, respectively (see Figure 3).

To find the difference of $T_{10\%}$, $T_{25\%}$, $C_{10\%}$, $C_{25\%}$ between these disciplines, we employ the Kruskal-Wallis H-test. Table 1 shows that Clinical, Pre-Clinical & Health, and Life Sciences are the first two disciplines that reach $C_{10\%}$ and $C_{25\%}$. Arts & Humanities is the discipline that needs more time to reach $C_{10\%}$ and $C_{25\%}$; however, as the number of articles in this discipline is very few, this may result in potential bias. As for the statistical differences, we find there is significant difference between Clinical, Pre-Clinical & Health, Life Sciences, and Physical Sciences for $T_{10\%}$ ($p < 0.001$). For $T_{10\%}$, there is not a significant difference between Engineering & Technology and Arts & Humanities ($p > 0.05$) nor between Social Sciences and Arts & Humanities ($p > 0.05$). There is not a significant difference between Arts & Humanities and other disciplines for $C_{10\%}$ and $C_{25\%}$ ($p > 0.05$), nor for Clinical, Pre-Clinical & Health and Life Sciences ($p > 0.05$). In addition, there is a significant difference between Clinical, Pre-Clinical & Health, Life Sciences, Physical Sciences, Engineering & Technology, and Social Sciences for $T_{25\%}$ ($p < 0.001$).

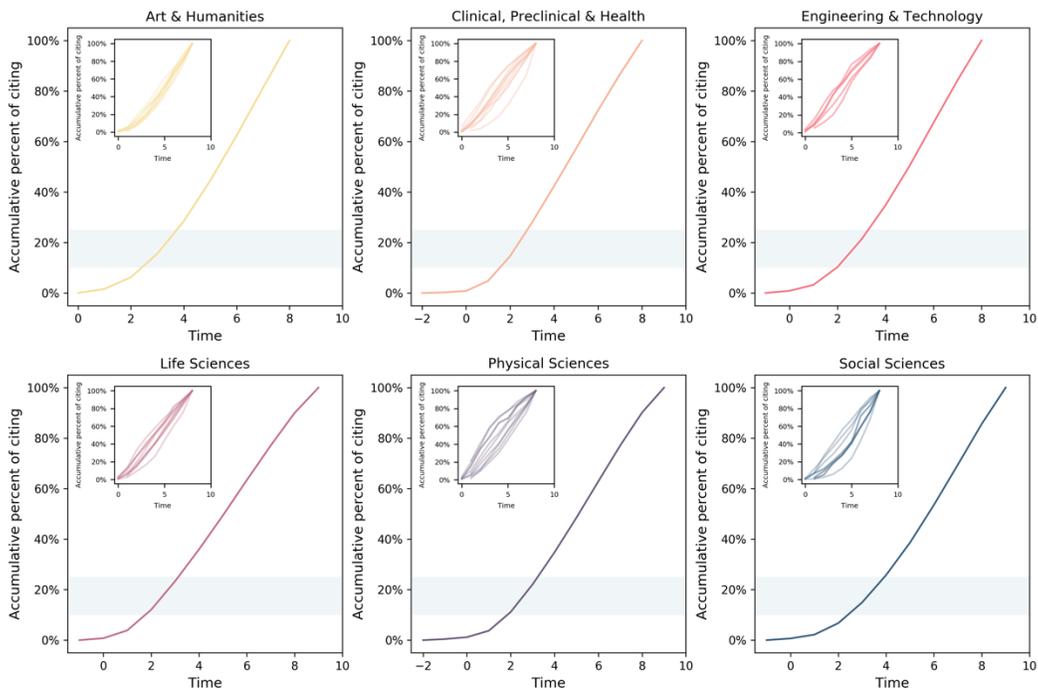

Fig. 3 Average S-shaped curve of highly cited papers in each discipline

Table 1. Time and citations at "take-off" stage between different disciplines

| Disciplines | Indicators (Mean value) | | | |
| --- | --- | --- | --- | --- |
| | $T_{10\%}$ | $C_{10\%}$ | $T_{25\%}$ | $C_{25\%}$ |
| Clinical, Pre-Clinical & Health | 1.61* | 53.25 | 2.83* | 104.56 |
| Life Sciences | 1.69* | 59.9 | 2.96* | 120.99 |
| Physical Sciences | 1.82* | 49.36 | 3.10* | 99.32 |
| Engineering & Technology | 2.02 | 40.6 | 3.35* | 84.77 |
| Social Sciences | 2.33 | 33.01 | 3.72* | 70.30 |
| Arts & Humanities | 2.60 | 65.8 | 3.70 | 116.40 |

*There is significant difference after Kruskal-Wallis H-test, $p = 0.05$.

*Knowledge broadness influences*

This section presents the time and citation counts from the knowledge broadness perspective. There are 2182 articles with the number of references less than or equal to 21, 3770 articles cite between 22 and 41 references, and the rest of 1975 papers cite greater than or equal to 42 references. Figure 4 shows that there seems to be no delayed recognition for papers that cite greater than or equal to 42 references than for the other two groups. We also find that highly cited papers with the number of references less than or equal to 21 need longer time to attract citations than those with more than 21 references ($p<0.001$). There is no statistical significance between papers with references equal to or more than 42 and those with references between 22 and 41 ($p>0.05$). In other words, highly cited papers with limited references usually take more time to reach 10% and 25% citations than papers that cite a moderate or large number of references (see Table 2.)

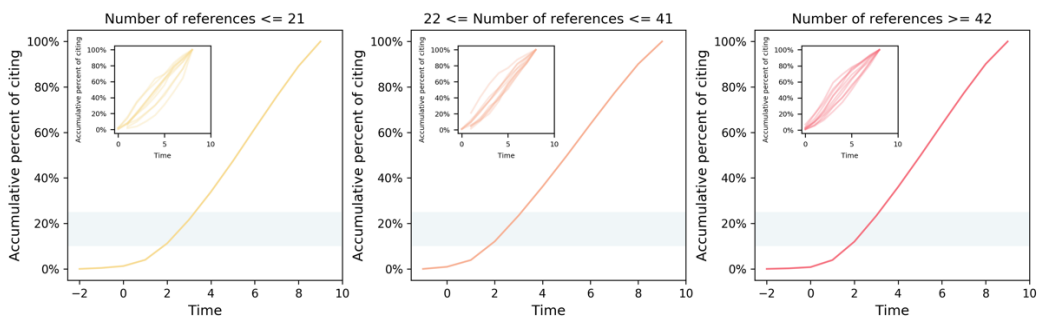

Fig. 4 Average S-shaped curve of highly cited papers between different knowledge broadness groups

Table 2. Time and citations at "take-off" stage between different knowledge broadness groups

| Knowledge broadness | Indicators (Mean value) | | | |
|---|---|---|---|---|
| | $T_{10\%}$ | $C_{10\%}$ | $T_{25\%}$ | $C_{25\%}$ |
| Number of references ≤21 | 1.92* | 48.83* | 3.21* | 99.01* |
| 22 ≤ Number of references ≤ 41 | 1.72 | 52.34 | 2.99 | 105.52 |
| Number of references ≥ 42 | 1.72 | 53.99 | 2.99 | 106.89 |

*There is significant difference after Kruskal-Wallis H-test, $p = 0.05$.

*Knowledge recency influences*

In this section, we explore the time and citations between different knowledge recency groups. There are 2086 publications where the Price index is equal to or less than 0.43, 3869 articles with the Price index greater than or equal to 0.44 and less than or equal to 0.71, and 1972 articles tend to cite more recently published articles with the Price index greater than or equal to 0.72.

Results show that the three different Price index groups have statistically significant differences between the four indicators ($p<0.001$). More specifically, the citation time of these articles that reach the 10% and 25% of total citations from longest to shortest sequence as: Price index less than or equal to 0.43, Price index between 0.44 and 0.71, and Price index equal to or greater than 0.72 ($p<0.001$). This also echoes the sequence of citation counts that reach the 10% and 25% of total citations from smallest to largest. This reflects that highly cited articles with more recent references easily attract citations in a shorter amount of time (see Figure 5 and Table 3).

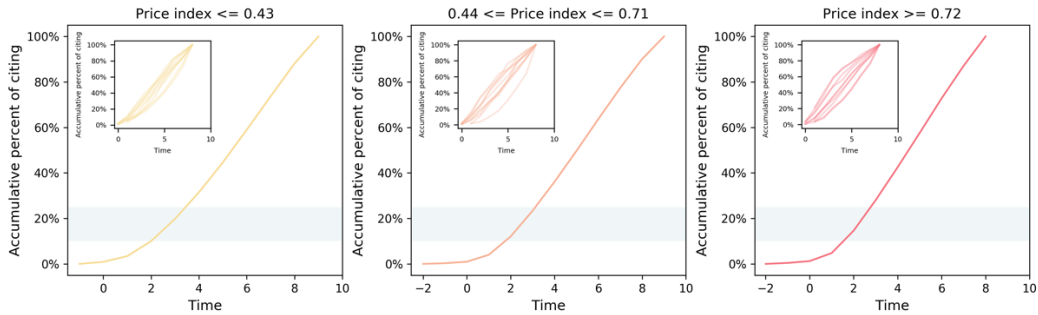

Fig. 5 Average S-shaped curve of highly cited papers between different knowledge recency groups

Table 3. Time and citations at "take-off" stage between different knowledge recency groups

| Knowledge recency | Indicators (Mean value) | | | |
| --- | --- | --- | --- | --- |
| | $T_{10\%}$* | $C_{10\%}$* | $T_{25\%}$* | $C_{25\%}$* |
| Price index ≤ 0.43 | 1.95 | 43.98 | 3.32 | 90.26 |
| 0.44 ≤ Price index ≤ 0.71 | 1.74 | 52.00 | 3.01 | 104.73 |
| Price index ≥ 0.72 | 1.66 | 59.62 | 2.85 | 117.37 |

*There is significant difference after Kruskal-Wallis H-test, $p = 0.05$.

*Multiple linear regression analysis*

The above sections have already analyzed the correlation between "take-off" time and factors like disciplines, knowledge broadness, and recency. However, such factors may be correlated with each other as well. In this section, multiple linear regression model is implemented to disentangle the contribution from each factor, and we set disciplines as a dummy variable and natural logarithm of number of references is used for model estimation. We apply ANOVA analysis to test the significance of our model. The results suggest that our regression model is significant ($p<0.001$) in explaining the "take-off" time in both $T_{10}$ and $T_{25}$ of highly cited papers (See Table 4 and Table 5). In addition, knowledge broadness, knowledge recency, and discipline are significant factors that influence the "take-off" time of highly cited papers ($p<0.05$). Specifically, knowledge broadness and recency have a negative effect on the "take-off" time, where unstandardized coefficient of knowledge recency is smaller than knowledge broadness. As for the disciplinary effect, our results show that Social Science and Art & Humanities have no significant difference ($p>0.05$) on the "take-off" time of highly cited papers. Engineering & Technology and Art & Humanities, Physical Science and Art & Humanities have no significant difference (p>0.05) on the "take-off" time when they reach the 25% of total citations."

Table 4 Regression results of "take-off" time at $T_{10\%}$

| Variables | $\beta$ (95% CI) | Std. Error | $t$ | $p$ |
|---|---|---|---|---|
| Constant | 2.969 (2.517~3.264) | 0.301 | 9.862 | 0.001 |
| No. of references | -0.001(-0.002~-3.989E-6) | 0.000 | -2.171 | 0.030 |
| Price_index | -0.407(-0.486~-0.328) | 0.039 | -10.546 | 0.001 |
| Life sciences | -1.036(-1.333~-0.588) | 0.301 | -3.440 | 0.001 |
| Physical sciences | -0.922(-1.207~-0.474) | 0.301 | -3.062 | 0.002 |
| Clinical, Preclinical & Health | -1.127(-1.419~-0.674) | 0.301 | -3.742 | 0.001 |
| Engineering & Technology | -0.688(-0.974~-0.228) | 0.301 | -2.284 | 0.022 |
| Social sciences | -0.420(-0.715~0.033) | 0.302 | -1.389 | 0.165 |
| Adjusted $R^2$ | | 0.1 | | |
| $p$ | | 0.001* | | |
| $N$ | | 7927 | | |

Table 5 Regression results of "take-off" time at $T_{25\%}$

| Variables | $\beta$ (95% CI) | Std. Error | $t$ | $p$ |
|---|---|---|---|---|
| Constant | 4.091(3.544~4.454) | 0.352 | 11.617 | 0.001 |
| No. of references | -0.001(-0.002~0.000) | 0.000 | -2.633 | 0.008 |
| Price_index | -0.714(-0.809~-0.623) | 0.045 | -15.822 | 0.001 |
| Life sciences | -0.706(-1.058~-0.158) | 0.352 | -2.004 | 0.045 |
| Physical sciences | -0.587(-0.934~-0.039) | 0.352 | -1.666 | 0.096 |
| Clinical, Preclinical & Health | -0.848(-1.197~-0.278) | 0.352 | -2.408 | 0.016 |
| Engineering & Technology | -0.279(-0.628~0.280) | 0.353 | -0.791 | 0.429 |
| Social sciences | -0.021(-0.380~0.530) | 0.353 | -0.059 | 0.953 |
| Adjusted $R^2$ | | 0.123 | | |
| $p$ | | 0.001* | | |
| $N$ | | 7927 | | |

**Discussions and conclusions**

Innovation diffusion theory is one of the oldest social science theories and is widely used in marketing. The main purpose of this study is to examine the threshold of these successfully disseminated creative ideas at their early stage through the lens of Rogers's innovation diffusion theory. Scientists give credit to colleagues by citing their work when used as part of their studies, and a highly cited paper has certainly and successfully disseminated scientific ideas to the scientific community. Therefore, we take highly cited papers as an example to explore the time span and citations at their "take-off" stage. This may provide insight into the threshold of time and number of citations at which research papers may quickly become highly cited.

We find that most (greater than 87%) of the highly cited papers attract an average of 50.46 and 102.93 citations within two years and four years of publication, respectively. These are the lower and upper threshold of "take-off" stage, based on the S-shaped diffusion model for highly cited research articles. In other words, a research article has the potential to become highly cited if it received at least 51 citations within two years, or 103 citations within four years. Even so, we should notice that this conclusion is not absolute; there are still a minimal number of papers showing the "delayed rise" pattern. Some of these papers accumulate only one citation within two years and four years of their publication but then suddenly receive a citation burst and become highly cited. In research evaluation, there is an essential tension between the needs of funders for timely assessment of funded research and the time it might take for research to reveal its impact; therefore, the application of these thresholds depends on this tension. Interestingly, our results also show that there are 0.66% (52) of articles that accumulate their first citations before publication, which means the "response time" of a minimal number of highly cited papers occurs even before their publication (Bornmann et al. 2017; Schubert and Glanzel 1986). We did not analyze the reasons of this situation, but one of the reasons may be a result of preprint. As for the increasing trend of preprint in recent years, this situation may be more universal for newly published highly cited papers.

This study also explores the influence of discipline, number of references, and Price index on the threshold of "take-off" period. We find that Clinical, Pre-Clinical & Health, and Life Sciences are the first two disciplines that reach the $C_{10\%}$ and $C_{25\%}$, and Arts & Humanities is the discipline that needs more time to reach their $C_{10\%}$ and $C_{25\%}$ than other disciplines. As for the number of references, we find that papers with number of references less than or equal to 21 need longer time to attract citations than those with references numbering more than 21. According to former studies (Ahlgren et al. 2018; Costas et al. 2012), number of references is related to the author's knowledge broadness; this study shows that researchers cite references that less than 21 may hamper their research in quickly attracting citations. According to Moed (H. F. Moed 1989), citing recently published articles in one's publication is an indication of "popularity" orientation. This study further shows that highly cited articles with more recent references easily and quickly attract citations. In other words, popularity oriented, highly cited papers attract citations easily. We also employed regression analysis to disentangle the contribution of each factors, and results show that knowledge broadness and recency have a negative effect on the "take-off" time, where knowledge recency has a strong negative influence on the "take-off" time than knowledge broadness. There are no disciplinary effects between Social Science and Art & Humanities on the "take-off" time of $T_{10}$, as well as Engineering & Technology and Art & Humanities, Physical Science and Art & Humanities on the "take-off" time of $T_{25}$.

Here, we emphasize our study's descriptive power and simplicity; it is simple and easy to implement and can be used by not highly-skilled scientometric users. This study is the first step to understanding the timespan and citations of the "take-off" stage for successfully disseminated creative ideas based on innovation diffusion theory, but there are some limitations. We focused on the data covered by WoS and do not analyze data that is not included in WoS. In addition, some of the references were deleted due to incomplete references, for example, references without author, title, accession number, or journal information. Moreover, this paper takes a 10-year citation window to show that the citation curve follows an S-shaped curve. Yet, there may not be such physical cap for citations, given enough time. Therefore, sleeping beauty papers, which peak much later than most articles are unlikely to be captured by a 10-year window. In our next study, we will try to test longer citation windows for papers, for example, in the 1990s.

**Acknowledgments**

This contribution is based upon work supported by The National Social Science Foundation of China under Grant No. 14BTQ030. We acknowledge the support of the Chinese Scholarship Council. We are grateful to Yi Bu for providing the dataset and Weiwei Gu, as well as the anonymous reviewer for helpful comments on earlier versions of this article.

**References**

Ahlgren, P., Colliander, C., & Sjögårde, P. (2018). Exploring the relation between referencing practices and citation impact: A large-scale study based on Web of Science data. *Journal of the Association for Information Science and Technology, 69*(5), 728-743. doi:10.1002/asi.23986

Aksnes, D., Rorstad, K., Piro, F., & Sivertsen, G. (2011). Are Female Researchers Less Cited? A Large-Scale Study of Norwegian Scientists. *Journal of the American Society for Information Science and Technology, 62*(4), 628-636. doi:10.1002/asi.2148610.1002/asi

Beaudry, C., & Larivière, V. (2016). Which gender gap? Factors affecting researchers' scientific impact in science and medicine. *Research Policy, 45*(9), 1790-1817. doi:10.1016/j.respol.2016.05.009

Bornmann, L. (2013). The problem of citation impact assessments for recent publication years in institutional evaluations. *Journal of Informetrics, 7*(3), 722-729. doi:10.1016/j.joi.2013.05.002

Bornmann, L., & Daniel, H. (2006). Selecting scientific excellence through committee peer review - A citation analysis of publications previously published to approval or rejection of post-doctoral research fellowship applicants. *Scientometrics, 68*(3), 427-440.

Bornmann, L., & Daniel, H. D. (2008). What do citation counts measure? A review of studies on citing behavior. *Journal of Documentation, 64*(1), 45-80. doi:10.1108/00220410810844150

Bornmann, L., Ye, A. Y., & Ye, F. Y. (2017). Sequence analysis of annually normalized citation counts: an empirical analysis based on the characteristic scores and scales (CSS) method. *Scientometrics, 113*(3), 1665-1680. doi:10.1007/s11192-017-2521-9

Bouabid, H. (2011). Revisiting citation aging: a model for citation distribution and life-cycle prediction. *Scientometrics, 88*(1), 199-211. doi:10.1007/s11192-011-0370-5

Bourdieu, P. (1991). The peculiar history of scientific reason. *Sociological Forum, 6*(1), 3-26. doi:10.1007/bf01112725

Boyack, K., & Klavans, R. (2014). Atypical combinations are confounded by disciplinary effects. *STI 2014 Leiden, 64*.

Callaham, M., Wears, R., & Weber, E. (2002). Journal prestige, publication bias, and other characteristics associated with citation of published studies in peer-reviewed journals. *Journal of the American Medical Association, 287*(21), 2847-2850.

Costas, R., van Leeuwen, T. N., & Bordons, M. (2012). Referencing patterns of individual researchers: Do top scientists rely on more extensive information sources? *Journal of the American Society for Information Science and Technology, 63*(12), 2433-2450. doi:10.1002/asi.22662

Didegah, F., & Thelwall, M. (2013a). Determinants of research citation impact in nanoscience and nanotechnology. *Journal of the American Society for Information Science and Technology, 64*(5), 1055-1064. doi:10.1002/asi.22806

Didegah, F., & Thelwall, M. (2013b). Which factors help authors produce the highest impact research? Collaboration,

journal and document properties. *Journal of Informetrics, 7*(4), 861-873. doi:10.1016/j.joi.2013.08.006

Egghe, L., Rao, I. K. R., & Rousseau, R. (1995). On the influence of production on utilization functions: Obsolescence or increased use? *Scientometrics, 34*(2), 285-315. doi:10.1007/bf02020425

Fortunato, S., Bergstrom, C. T., Borner, K., Evans, J. A., Helbing, D., Milojevic, S., Barabasi, A. L. (2018). Science of science. *Science, 359*(6379). doi:10.1126/science.aao0185

Gosnell, C. F. (1943). *The rate of obsolescence in college library book collections / as determined by an analysis of three select lists of books for college libraries* (Thesis edition ed.). New York: New York University.

Haslam, N., Ban, L., Kaufmann, L., Loughnan, S., Peters, K., et al. (2008). What makes an article influential? Predicting impact in social and personality psychology. *Scientometrics, 76*(1), 169-185. doi:10.1007/s11192-007-1892-8

Hu, Z., & Wu, Y. (2018). A probe into causes of non-citation based on survey data. *Social science information sur les sciences sociales, 57*(1), 139-151.

King, J. (1987). A review of bibliometric and other science indicators and their role in research evaluation. *Journal of Information Science, 13*(5), 261-276.

Kong, X., Jiang, H., Wang, W., Bekele, T. M., Xu, Z., et al. (2017). Exploring dynamic research interest and academic influence for scientific collaborator recommendation. *Scientometrics, 113*(1), 369-385. doi:10.1007/s11192-017-2485-9

Lee, Y.-N., Walsh, J. P., & Wang, J. (2015). Creativity in scientific teams: Unpacking novelty and impact. *Research Policy, 44*(3), 684-697. doi:10.1016/j.respol.2014.10.007

Li, J., & Ye, F. Y. (2016). Distinguishing sleeping beauties in science. *Scientometrics, 108*(2), 821-828.

Lillquist, E., & Green, S. (2010). The discipline dependence of citation statistics. *Scientometrics, 84*(3), 749-762. doi:10.1007/s11192-010-0162-3

Merton, R. K. (1961). Singletons and Multiples in Scientific Discovery: A Chapter in the Sociology of Science. *Proceedings of the American Philosophical Society, 105*(5), 470-486.

Merton, R. K. (1968). The Matthew Effect in Science: The reward and communication systems of science are considered. *Science, 159*(3810), 56-63. doi:10.1126/science.159.3810.56

Min, C., Ding, Y., Li, J., Bu, Y., Pei, L., et al. (2018). Innovation or immitation: the diffusion of citations. *Journal of the Association for Information Science and Technology, 69*(10):1271-1282. doi.org/10.1002/asi.24047

Moed, H. F. (1989). Bibliometric measurement of research performance and Price's theory of differences among the sciences. *Scientometrics, 15*(5-6), 473-483. doi:10.1007/bf02017066

Moed, H. F., Burger, W. J. M., Frankfort, J. G., & Raan, A. F. J. V. (1985). The use of bibliometric data for the measurement of university research performance. *Research Policy, 14*(3), 131-149. doi:doi.org/10.1016/0048-7333(85)90012-5

Mukherjee, S., Romero, D. M., Jones, B., & Uzzi, B. (2017). The nearly universal link between the age of past knowledge and tomorrow's breakthroughs in science and technology: The hotspot. *Sci Adv, 3*(4), e1601315. doi:10.1126/sciadv.1601315

Onodera, N., & Yoshikane, F. (2015). Factors affecting citation rates of research articles. *Journal of the Association for Information Science and Technology, 66*(4), 739-764. doi:10.1002/asi.23209

Pobiedina, N., & Ichise, R. (2016). Citation count prediction as a link prediction problem. *Applied Intelligence, 44*(2), 252-268. doi:10.1007/s10489-015-0657-y

Rinia, E., van Leeuwen, T., Eppo, B., Hendrik, V. V., & van Raan, A. F. J. (2001). Citation delay in interdisciplinary

knowledge exchange. *Scientometrics, 51*(1), 293-309.

Rogers, E. M. (1995). *Diffusion of innovations* (Fourth Edition ed.). New York: The Free Press.

Roldan-Valadez, E., & Rios, C. (2015). Alternative bibliometrics from impact factor improved the esteem of a journal in a 2-year-ahead annual-citation calculation: multivariate analysis of gastroenterology and hepatology journals. *Eur J Gastroenterol Hepatol, 27*(2), 115-122. doi:10.1097/MEG.0000000000000253

Roth, C., Wu, J., & Lozano, S. (2012). Assessing impact and quality from local dynamics of citation networks. *Journal of Informetrics, 6*(1), 111-120. doi:10.1016/j.joi.2011.08.005

Schubert, A., & Glanzel, W. (1986). Mean response time - A new indicator of journal citation speed with application to physics journals. Czecholovak journal of physics*, 36*(1), 121-125.

Simonton, D. K. (1997). Creative productivity: A predictive and explanatory model of career trajectories and landmarks. *Psychological Review, 104*(1), 66-89. doi:10.1037/0033-295x.104.1.66

Tahamtan, I., Safipour Afshar, A., & Ahamdzadeh, K. (2016). Factors affecting number of citations: a comprehensive review of the literature. *Scientometrics, 107*(3), 1195-1225. doi:10.1007/s11192-016-1889-2

Uzzi, B., Mukherjee, S., Stringer, M., & Jones, B. (2013). Atypical combinations and scientific impact. *Science, 342*(6157), 468-472. doi:10.1126/science.1240474

van Raan, A. F. J. (2004). Sleeping Beauties in science. *Scientometrics, 59*(3), 467-472.

Vanclay, J. K. (2013). Factors affecting citation rates in environmental science. *Journal of Informetrics, 7*(2), 265-271. doi:10.1016/j.joi.2012.11.009

Wang, D., Song, C., & Barabasi, A. L. (2013). Quantifying long-term scientific impact. *Science, 342*(6154), 127-132. doi:10.1126/science.1237825

Wang, J. (2013). Citation time window choice for research impact evaluation. *Scientometrics, 94*(3), 851-872. doi:10.1007/s11192-012-0775-9

Yan, R., Tang, J., Liu, X., Shan, D., & Li, X. (2011a). *Citation count prediction: Learning to estimate future citations for literature.* Paper presented at the CIKM '11 Proceedings of the 20th ACM international conference on Information and knowledge management, Glasgow, Scotland, UK

Yan, R., Tang, J., Liu, X., Shan, D., & Li, X. (2011b). *Citation count prediction: learning to estimate future citations for literature.* Paper presented at the Proceedings of the 20th ACM international conference on Information and knowledge management.

Yu, T., Yu, G., Li, P.-Y., & Wang, L. (2014). Citation impact prediction for scientific papers using stepwise regression analysis. *Scientometrics, 101*(2), 1233-1252. doi:10.1007/s11192-014-1279-6

Zhou, W., Gu, J., & Jia, Y. (2018). h-Index-based link prediction methods in citation network. *Scientometrics, 117*(1), 381-390. doi:10.1007/s11192-018-2867-7